\documentclass[12pt,preprint]{aastex}
\pdfoutput=1

\newcommand{\rhb}{R$\rm _{F487N}$}

\newcommand{\Ha}{H$\alpha$}

\newcommand{\Hb}{H$\beta$}

\newcommand{\htwo}{H$_{2}$}

\newcommand{\cmq}{cm$^{-3}$}

\newcommand{\oiii}{[O~III]}

\newcommand{\nii}{[N~II]}
\newcommand{\sii}{[S~II]}

\newcommand{\Ne}{n$\rm_{e}$}
\newcommand{\Te}{T$\rm_{e}$}
\newcommand{\cHb}{c$_{H\beta}$}

\newcommand{\Rc}{\rm R$\rm _{F547M}$}

\newcommand{\ngc}{the Ring Nebula}
\newcommand{\hezone}{He$^{++}$+H$^{+}$}

\newcommand{\nzone}{He$\rm ^{o}$+H$^{+}$}

\newcommand{\tsqa}{$t^{2}_A$}

\slugcomment{To appear in the Astronomical Journal}

\shorttitle{NGC 6720 Physical Conditions}
\shortauthors{O'Dell}

\begin{document}

\title{Studies of NGC 6720 with Calibrated HST WFC3 Emission-Line Filter Images--II:Physical Conditions\
\footnote{
Based on observations with the NASA/ESA Hubble Space Telescope,
obtained at the Space Telescope Science Institute, which is operated by
the Association of Universities for Research in Astronomy, Inc., under
NASA Contract No. NAS 5-26555.}
\footnote{Based on observations at the San Pedro Martir Observatory operated by the  Universidad Nacional Aut\'onoma de M\'exico.}}

\author{C. R. O'Dell}
\affil{Department of Physics and Astronomy, Vanderbilt University, Box 1807-B, Nashville, TN 37235}

\author{G. J. Ferland}
\affil{Department of Physics and Astronomy, University of Kentucky, Lexington, KY 40506}

\author{W. J. Henney}
\affil{Centro de Radioastronom\'{\i}a y Astrof\'{\i}sica, Universidad Nacional Aut\'onoma de M\'exico, Apartado Postal 3-72,
58090 Morelia, Michaoac\'an, M\'exico}

\and

\author{M. Peimbert}
\affil{Instituto de Astronomia, Universidad Nacional Aut\'onoma de M\'exico, Apdo, Postal 70-264, 04510 M\'exico D. F., M\'exico}

\email{cr.odell@vanderbilt.edu}

\begin{abstract}

We have performed a detailed analysis of the electron temperature and density in the \ngc\ using the calibrated HST WFC3 images described in the preceding paper. The electron temperature (\Te) determined from \nii\ and \oiii\ rises slightly and monotonically towards the central star. The observed equivalent width (EW) in the central region indicates that \Te\ rises as high as 13000 K. In contrast, the low EW's in the outer regions are largely due to scattered diffuse Galactic radiation by dust.  
The images allowed determination of unprecedented small scale variations in \Te. These variations indicate that the mean square area temperature fluctuations are significantly higher than expected from simple photoionization. The power producing these fluctuations occurs at scales of less than 3.5$\times$10$^{15}$ cm. This scale length provides a strong restriction on the mechanism causing the large t$^2$ values observed. 

\end{abstract}

\section{Background and Introduction}
\label{sec:intro}

In the preceding paper we have presented calibrated emission-line images of \ngc\ obtained with the Hubble Space Telescope's WFC3. These were combined with data from ground-based telescope high velocity resolution spectra to develop a detailed 3-D model of this prototypical planetary nebula. In this paper we present an analysis of the physical conditions.

The physical conditions in NGC~6720 have been the subject of numerous studies. In terms of the spatial coverage, the most complete investigation is that of Lame \&\ Pogge (1994), where a Fabry-Perot spectrophotometer was used to map the nebula in  the \Ha, \Hb, \nii\ 658.3 nm + 575.5 nm, \sii\ 671.6 nm + 673.1 nm, [O~I] 630.0 nm, \oiii\ 500.7 nm, and He~II 468.6 nm lines. Our present study can be viewed as an extension of the Lame \&\ Pogge (1994) and  Guerrero et al. (1997, henceforth GMC) investigations, but we have also observed the 436.3 nm line of \oiii\  and with more than 20x improvement in spatial resolution over the ground-based studies, although with the sacrifice of coverage of about one-half of the Main Ring of the nebula. Among the numerous slit spectroscopy studies the most recent and comprehensive high resolution investigation was that of O'Dell et al. (2007b, henceforth OSH), and the lower resolution multi-line studies of Garnett \&\ Dinerstein (2001) and OHS have also appeared. In the preceeding paper we develop and present the most up-to-date 3-D model for the nebula.


In the first part of this paper we characterize the extinction (\S\ \ref{sec:Ext}), the variation in the electron temperature (\Te) (\S\ \ref{sec:temps}) and electron density (\Ne) (\S\ \ref{sec:densities}) in a quadrant of the object, derive the Equivalent Widths (EW) of the underlying continuum (\S\ \ref{sec:EWs}), and then evaluate the magnitude and importance of small-scale \Te\ variations (\S\ \ref{sec:variations}). 

\section{Analysis of \ngc\  Calibrated Images}
\label{sec:Anal}

\subsection{Derived Extinction Values}
\label{sec:Ext}

\begin{figure}
\epsscale{1.0}
\plotone{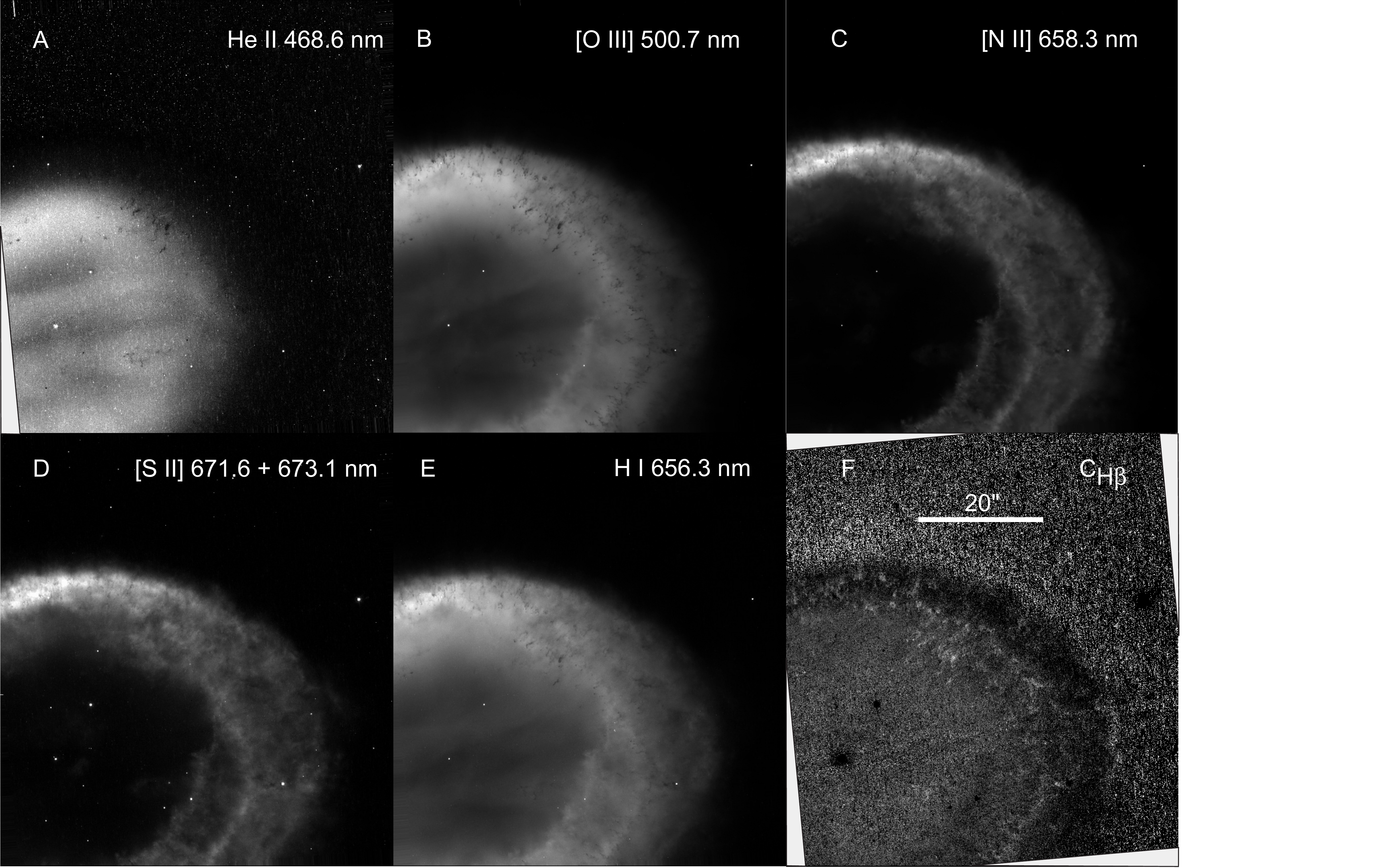}
\caption{This mosaic shows the 62.9\arcsec $\times$69.4\arcsec\ section of the NGC~6720 images common to both the full field and quadrant filters.  Panels A through E show the emission-line linear images corrected for contamination, if any, for the underlying continuum and non-targeted emission-lines, normalized to the maximum features being white.  Panel F shows the extinction index \cHb\ derived from the surface brightness in \Ha\ and \Hb\ with the scale indicated in Figure \ref{fig:f2}. The parallel faint linear features passing through the Central Star in the \cHb\ image are residual ``scars'' caused by small photometric errors near the edges of the two CCD detectors of the WFC3-VIS. The pointing was changed between two otherwise identical exposures to allow coverage of the gap between the detectors.The vertical axis is pointed towards PA=330\arcdeg.
\label{fig:f1}}
\end{figure}

We have used the calibrated emission-line images (Figure~\ref{fig:f1}, panels A-E) to determine the spatial distribution of the extinction-corrected emission. These images were then used to determine the ionization structure within the quadrant of the Main Ring of the nebula and the electron temperature and density distributions.

\begin{figure}
\epsscale{1.0}
\plotone{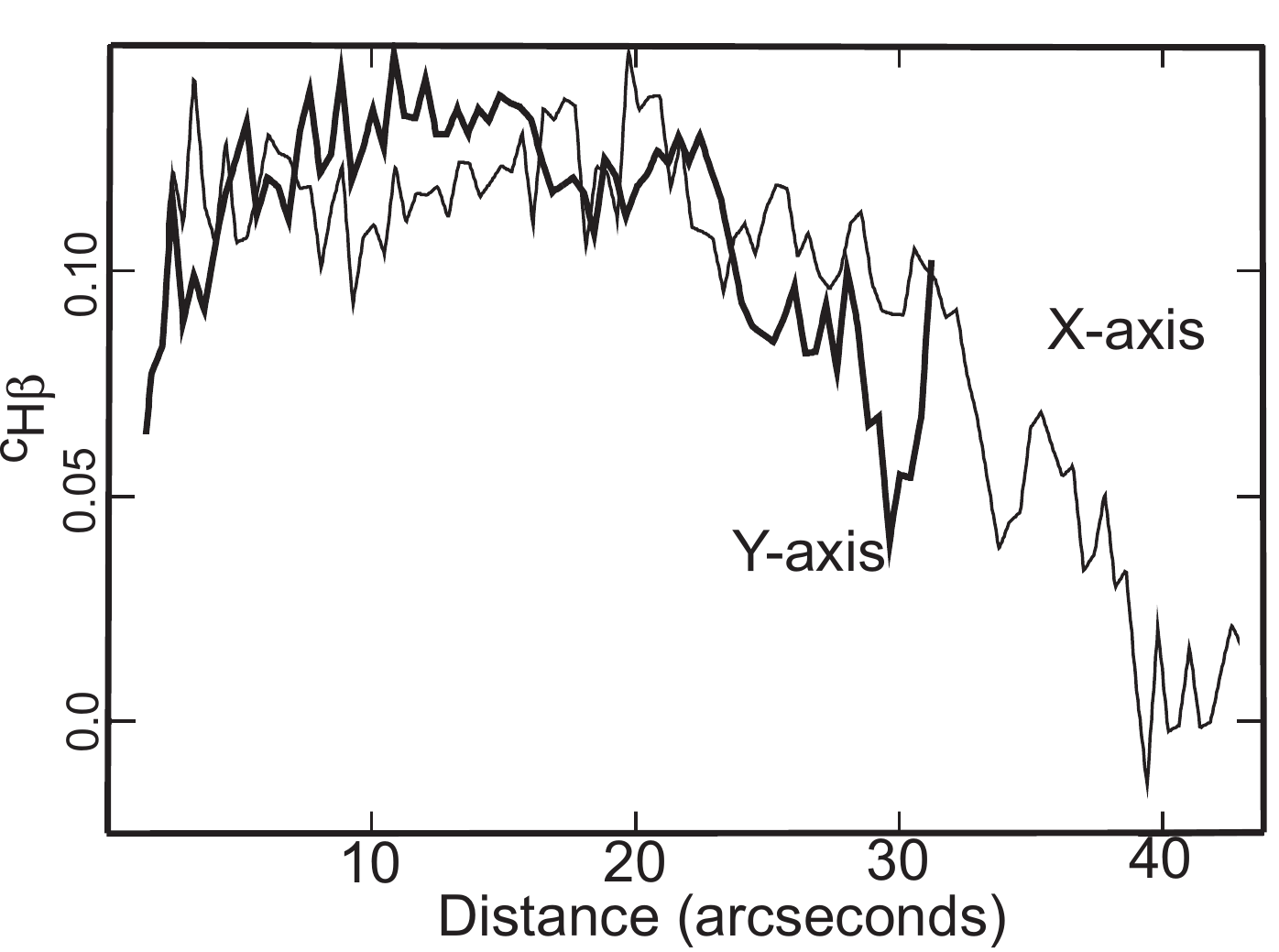}
\caption{The extinction factor \cHb\ is shown as derived from the observed \Ha/\Hb\ ratio and assuming an intrinsic ratio of 2.87, which applies to a low density gas at \Te\ = 10000 K.  The X-axis lies along PA = 240\arcdeg\ and the Y axis along PA = 330\arcdeg. The sample was 4\arcsec\ wide and the data have been averaged along bins of ten pixels, corresponding to 0.4\arcsec. The distance shown is from the position of the Central Star. The data closest to that star have been edited out because of contamination, thus causing a central gap or in later similar images, artificial values near the Central Star.
\label{fig:f2}}
\end{figure}

The extinction correction was determined from the \Ha/\Hb\ line flux ratio. We adopted an intrinsic ratio 2.87 based on the predictions for a low density gas at an electron temperature of 10000 K \citep{ost06} and the extinction curve for a ratio of total to selective extinction of 3.1 from Table 7.1 of Osterbrock \&\ Ferland (2006). The results are best expressed in terms of \cHb, the logarithm of the extinction (base 10) at 486.1 nm and the results are shown in Figure~\ref{fig:f1} (panel F). Profiles of \cHb\ along the X-axis (PA = 240\arcdeg, the major semi-axis of the elliptical Main Ring) and the Y-axis (PA = 330\arcdeg, the minor semi-axis) are shown in Figure~\ref{fig:f2}. We see there that \cHb\ has a radial dependence, being about 0.12 in the central core out to the the inner Main Ring, then dropping near the ionization boundary. We will see in \S\ \ref{sec:temps} that there is not a significant variation in \Te\ with distance from the Central Star except for a slight increase within 20\arcsec\  of the Central Star. Since 20\arcsec\ is about where \cHb\ begins to drop, it is worth considering if the calculated \cHb\ is due to a decrease in \Te. Osterbrock \&\ Ferland (2006) Table 4.4 shows that the \Ha/\Hb\ ratio decreases by 4 \%\ in going from 10000 K to 20000 K. The indicated rise in \Te\ towards the center would produce a decrease in the \Ha/\Hb/ ratio of less than 2 \%, which would mean that we have under-estimated \cHb\ by  less than 0.025 and the true \cHb\ in the core would be even larger than is shown in Figure~\ref{fig:f2}. To explain away the outwards decrease in \cHb\ would require \Te\ to increase within the Main Ring, but the derived electron temperatures do not show this and the outward decrease is likely to be real.

The average \cHb\ is even higher in regions containing the knots seen in silhouette \citep{ode02}, which are not numerous in the two samples used in making Figure~\ref{fig:f2}. The inner Main Ring extinction of \cHb\ = 0.12 is similar to 0.15 found by Garnett \&\ Dinerstein (2001) along a line with PA = 91\arcdeg, and GMC, which found \cHb\ = 0.14. The Fabry-Perot  study \citep{lam94} does not give extinction values explicitly, but cite an overall \Ha/\Hb\ line flux ratio of 3.74$\pm$0.22 and 3.67 for the inner 15\arcsec. These ratios would correspond to \cHb\ values of 0.33 and 0.31 respectively. Since these inferred \cHb\ values are much higher than those from the several other studies, it appears likely that the flux calibration is not accurate.

\subsection{Electron Temperatures}
\label{sec:temps}

We derived the electron temperature for both the \oiii\ and \nii\ emitting regions using equations 5.4 and 5.5 of Osterbrock \&\ Ferland (2006) in their low density approximations. The nebular to auroral line ratios were obtained from our emission-line images as described in Appendix A.
The results are shown in Figure~\ref{fig:f3} for a 4.0\arcsec\ wide sample along our X axis (PA = 240\arcdeg) and in Figure~\ref{fig:f4} for a similar sample along our Y axis (PA = 330\arcdeg). 
The range of distances was limited to those where the relevant surface brightnesses were high. \nii\ emission arises from only in the Main Ring, giving a nearly constant temperature of about 10000 K, with indication of a slight increase at the inner boundary of the Main Ring. \oiii\ is strong from the center of the nebula out to close to the outer boundary of the Main Ring.
The \oiii\ temperature is very similar to the \nii\  temperature where both are available (essentially in the Main Ring).  However, the trend suggested by the \nii\ temperatures is confirmed and extended by the \oiii\ temperatures. 
There is a clear rise in the \oiii\ temperature to a central region value of about 11700 K. This inner-region is counter-intuitive since photoionization theory indicates that the electron temperature should rise with increasing distance from the ionizing star due to the effects of radiation hardening. 
The resolution of this riddle is that either processes affecting the heating become more important at the lower densities that apply to the central region or that the central \oiii\ emission arises from material physically further away from the Central Star and is only seen in projection at smaller angular separation. 

\begin{figure}
\epsscale{1.0}
\plotone{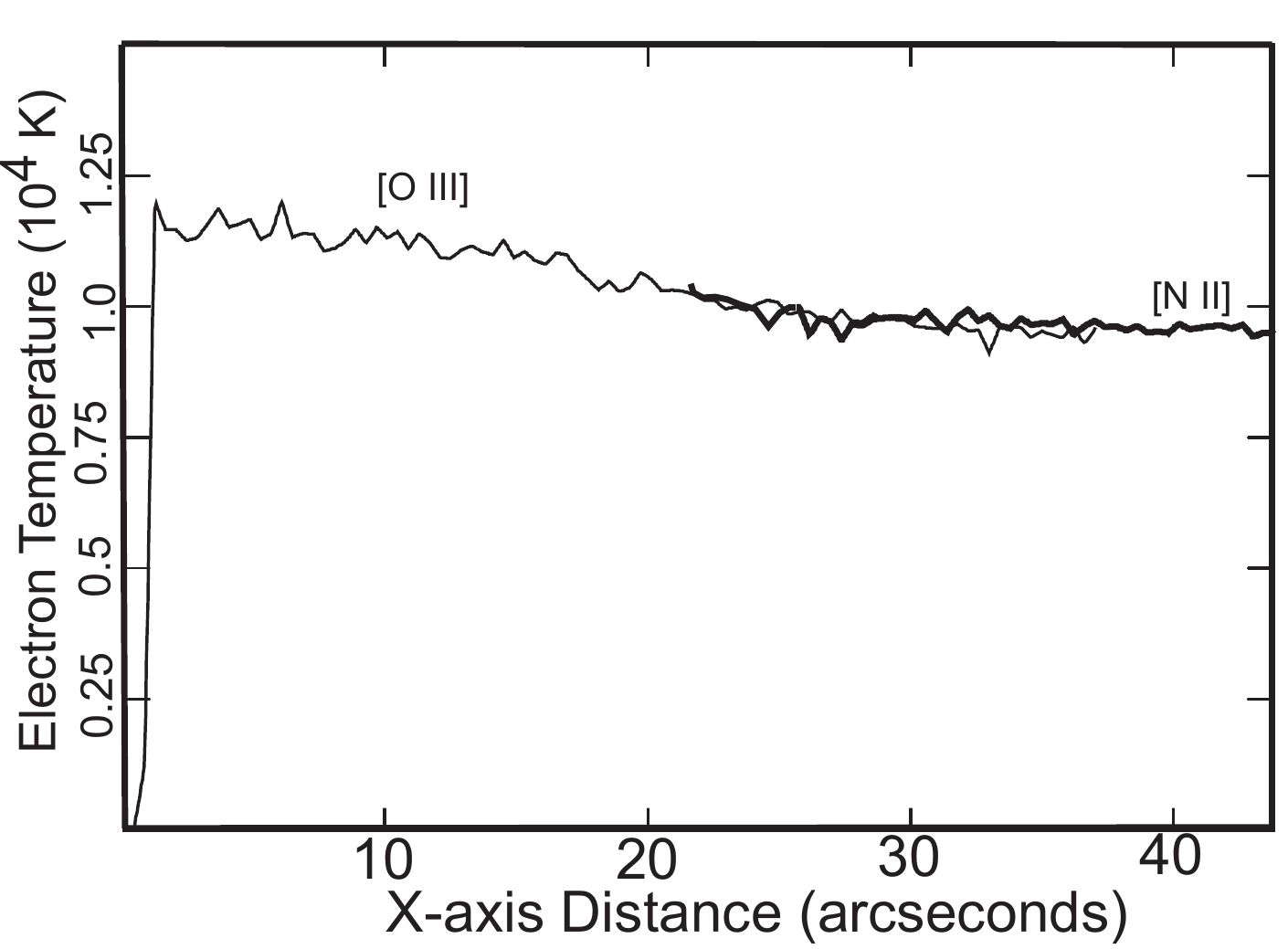}
\caption{The \Te\ values derived from  both the \nii\ and \oiii\ lines are shown for the X-axis profile. The range of results for the two ions was constrained to those where the surface brightness of the nebular transitions were the highest.
\label{fig:f3}}
\end{figure}

\begin{figure}
\epsscale{1.0}
\plotone{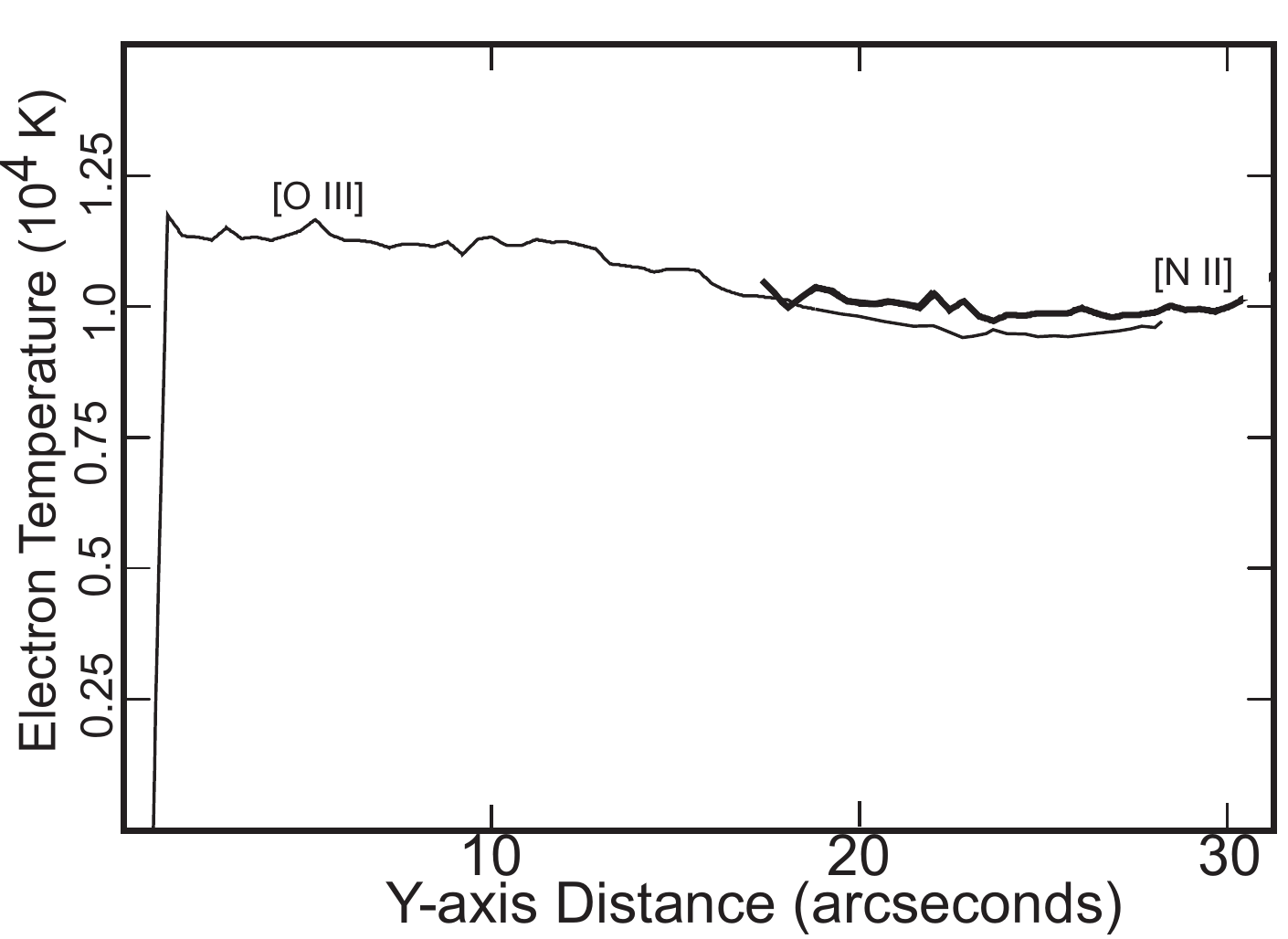}
\caption{Like Figure~\ref{fig:f3} except for the Y-axis.
\label{fig:f4}}
\end{figure}

Electron temperatures have been derived many times for \ngc, largely in connection with efforts to determine elemental abundances.  The observational line ratios used in their determination have probably improved over the decades and at the same time there have been changes in the adopted values of the atomic coefficients such as collision strengths and transition probabilities.  Limiting ourselves to the most recent work, we note that GMC find average values in the Main Ring of 10800 K from \nii\ and 11700 K from \oiii, with central region temperatures of 12600 K from \oiii. Garnett \&\ Dinerstein (2001) found the \oiii\ temperature to be 10000 K in the Main Ring, rising to about 12000 in the central region.  Lame \&\ Pogge (1994) did not detect a variation in their \nii\ temperature and found a global average of 9643$\pm$230 K. Liu et al. (2004) found \Te(\oiii) = 10600 K and \Te(\nii) = 10200 K. Our temperatures are within the range of these earlier spectroscopic results and we confirm the presence of a central increase (which is also seen in the Kupferman (1983) study).

\subsection{Electron Densities}
\label{sec:densities}

\begin{figure}
\epsscale{1.0}
\plotone{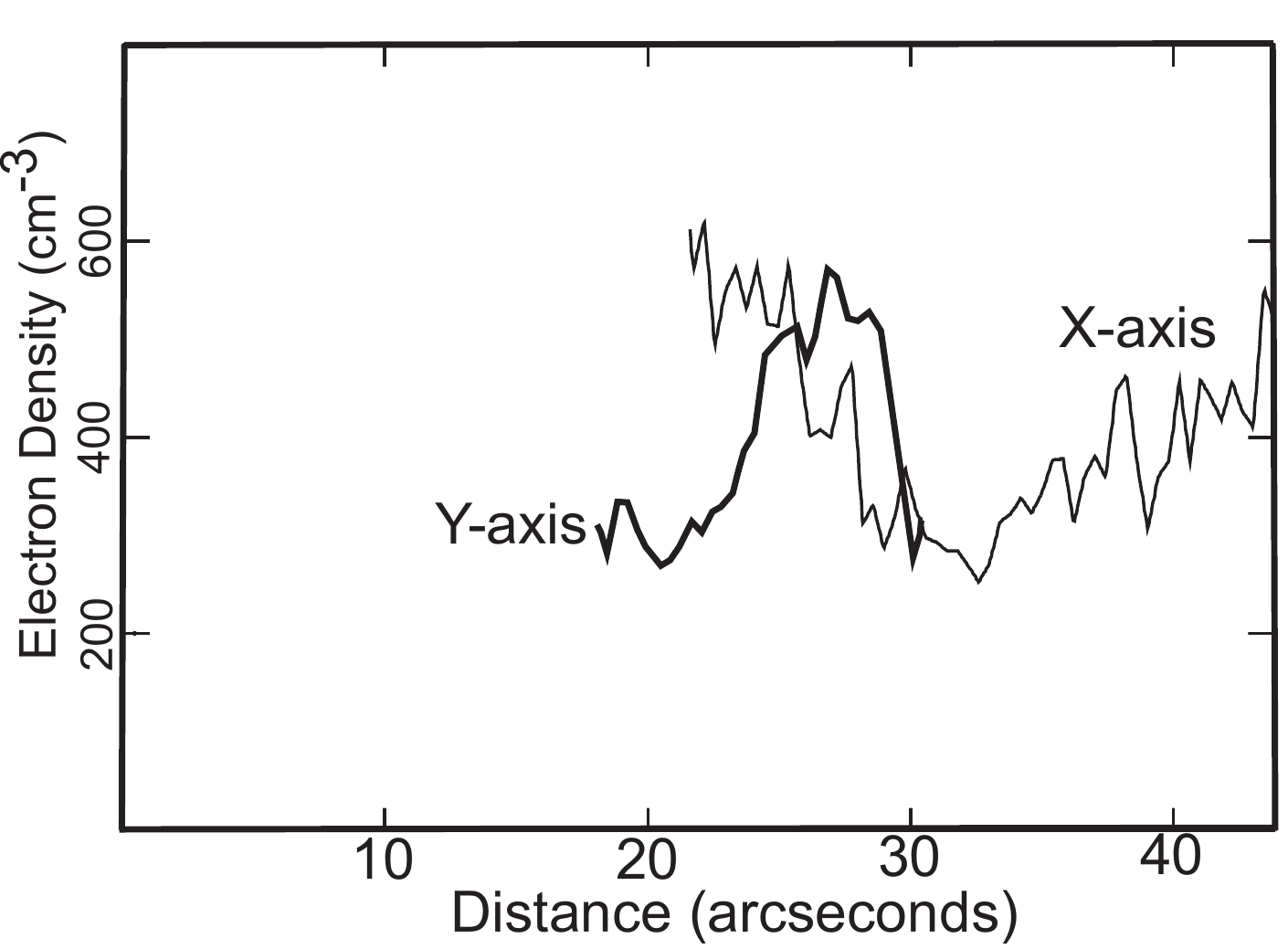}
\caption{The electron densities derived from the \sii\ 671.6 nm and 673.1 nm lines for an assumed electron temperature of 10000 K are shown form both the X-axis  (light line) and Y-axis (heavy line) profiles.
\label{fig:f5}}
\end{figure}

The densities in the \sii\ emitting regions were determined by adopting the relation log~\Ne\  = 4.705 -1.9875 S(671.6)/S(673.1) from Figure 5.8 of Osterbrock \&\ Ferland (2006), which is quite accurate over the line ratio range of 1.35 to 0.65, which corresponds to densities of about 100 \cmq\  to 3000 \cmq. The results are shown in Figure~\ref{fig:f5} for the region of high \sii\ surface brightness. We see that in the \sii\ bright regions of the Main Ring that peak values of about 550 \cmq\ are found but regions about 350 \cmq\ are present. Aller et al. (1976) measured \sii\ in three points in the Main Ring and when their observed line ratios are interpreted with the up-to-date atomic parameters in Osterbrock \&\ Ferland (2006) yield densities of 600, 360, and 170 \cmq. Lame \&\ Pogge (1994) found about 700 \cmq\ for the brightest part of the minor axis  (where we find 570 \cmq)and 500 \cmq\ for the rest of the nebula.  The GMC densities are higher than our own in that they found 615 \cmq\ for their OR sample (radially averaged over 25\arcsec -35\arcsec, which is not the same region we have sampled, even though Figure~\ref{fig:f5} gives results for those distances.).  Wang et al. (2004) determined densities in the Main Ring from multiple doublets and found 525 \cmq\ from the \sii\ doublet.  Given the large local variations in the true densities, we find good agreement in the results from the several studies and our derived densities.

\subsection{Equivalent Widths}
\label{sec:EWs}

The continuum radiation in \ngc\ is unusually strong. Since the continuum primarily arises from ionized gas processes, it is customary to express the strength relative to the the \Hb\ emission-line, the Equivalent Width (EW). EW is calculated as S(\Hb )/S$_{\lambda}$, where S(\Hb ) is the \Hb\ surface brightness in energy units (e.g. ergs) and S$_{\lambda}$ is the surface brightness brightness of the continuum in energy units per wavelength interval (e.g. ergs/\AA) at the same wavelength with the result being expressed in wavelength units (\AA\ in our case). EW is how wide a sample of the continuum is required to produce as much energy as the reference emission-line. This means that increasing strength of the continuum relative to \Hb\ produces a smaller EW, an awkward but traditional measure.

The several atomic processes producing the atomic continuum are detailed in Osterbrock \&\ Ferland 2006 and these include for hydrogen  free-bound, free-free, and two-photon emission and the same processes for contributions from both stages of ionized helium (with the two-photon component being relatively unimportant near the wavelength of \Hb). The hydrogen 2S level from which two-photon emission arises can be depopulated by collisions or radiation, with the critical density at which the two processes occur equally being 15500 \cmq. Since we show in \S\ \ref{sec:densities} that the highest
densities are about 500 \cmq, collisional quenching of two-photon emission is unimportant.  For a pure hydrogen gas at a 10000 K, EW would be 1400 \AA. The helium contribution to the continuum is primarily from doubly ionized helium, with the emissivity per ion being about seven times that of hydrogen.  If the relative number abundance of helium to hydrogen is 0.12 and the helium is 70 \%\ doubly ionized \citep{gar01}, the expected EW drops to 935 \AA.  For these same conditions the EW drops with increasing electron temperature, being 835 \AA\ at 12000 K,  770 \AA\ at 14000 K,  and 750 \AA\ at 16000 K. 

\begin{figure}
\epsscale{1.0}
\plotone{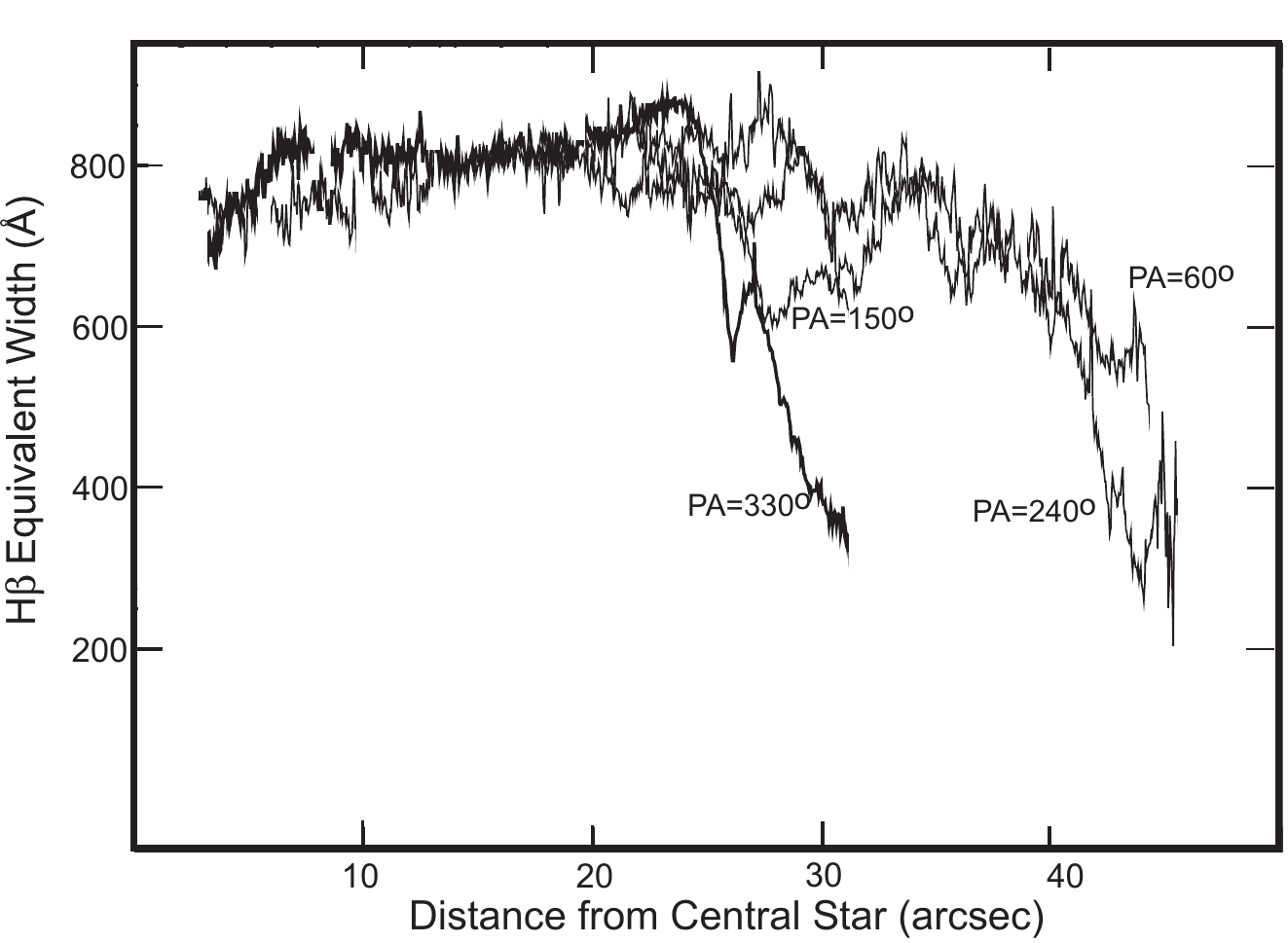}
\caption{The Equivalent Width of the \Hb\ emission line and its underlying continuum are shown along the PA values 60\arcdeg, 150\arcdeg, 240\arcdeg, and 330\arcdeg\ as determined from the WFC3 F547M and F487N images.  Decreasing Equivalent Width values indicate relatively stronger continuum values. Most of the regions contaminated by starlight have been edited out.
\label{fig:f6}}
\end{figure}

\begin{figure}
\epsscale{1.0}
\plotone{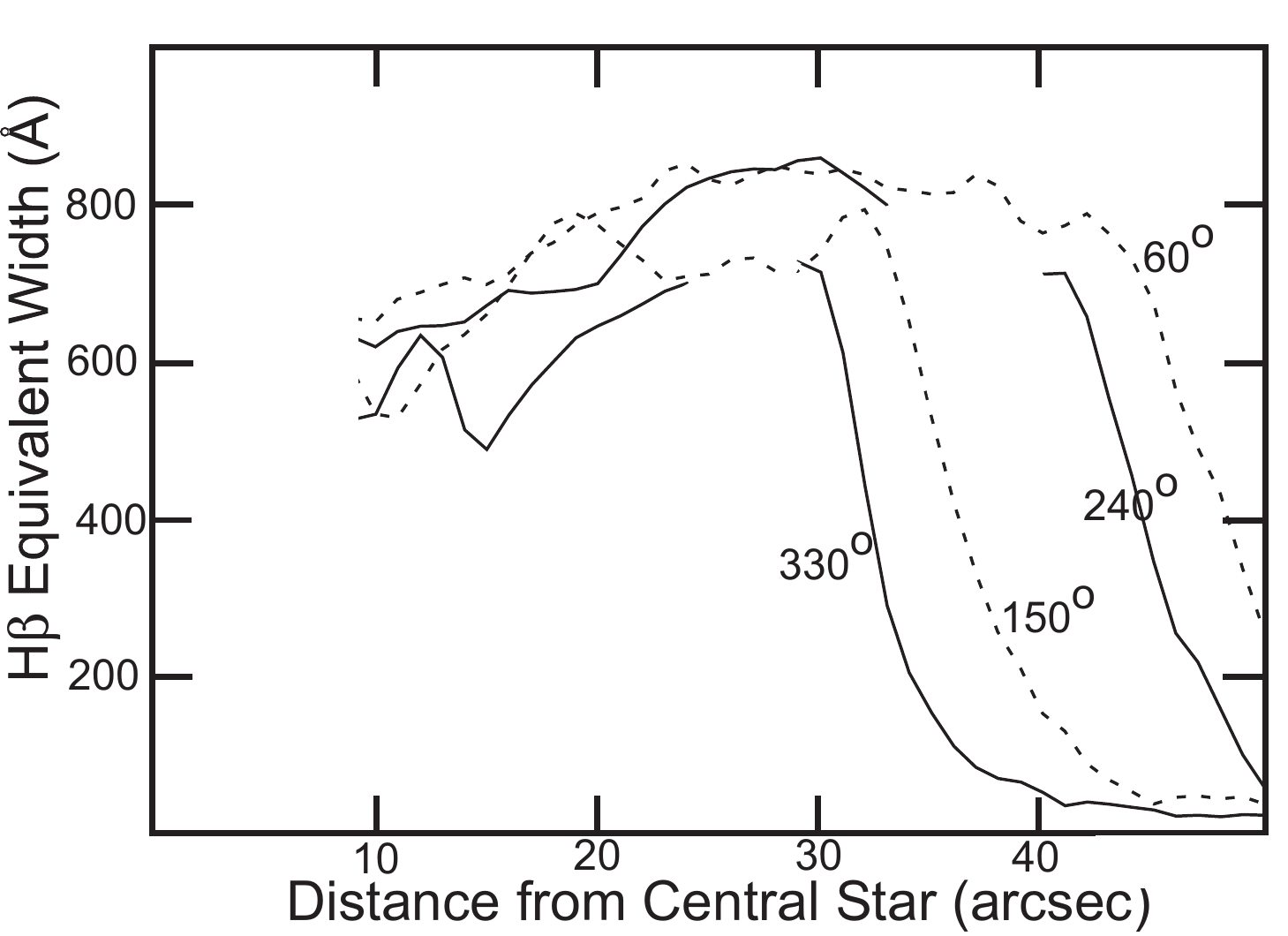}
\caption{Like Figure~\ref{fig:f6}, except that the EW values are derived from the ground-based spectra used for calibration. Different PA values are indicated. The regions contaminated by starlight have been edited out.
\label{fig:f7}}
\end{figure}

We have determined the EW at \Hb\ using the properties of the WFC3 filters and adopting a constant flux ratio (f$_{\lambda 4861}$/f$_{\lambda 5470}$ = 1.0), finding EW(\AA) = 690$\times$(\rhb/\Rc) -62. Figure~\ref{fig:f6} shows the derived EW along the primary axes of the nebula. If the flux ratio is 0.90$\pm$0.14, as determined from the spectra, then the derived EW values would be about 10\%\ larger. The observed WFC3 EW values indicate a central region temperature of about 13000 K. As an independent check on the unexpectedly low value of EW, we derived the EW from the spectra and show the results in Figure~\ref{fig:f7}. Although derived with rigor, there may be a small overestimate of the continuum strength (hence smaller EW) due to broad instrumental scattering of the very strong \oiii\ 495.9 nm line.  We see good agreement from these two methods, with the spectroscopic results being slightly lower, probably because of the instrumental scattered light. In this derivation of EW we noted that there was an almost constant background signal in both the F487N and F547M signals, being 2 \%\ of the Main Ring signal in F487N and about 10 \%\ for the F547M signal. These backgrounds were subtracted before the EW was calculated. 

\begin{figure}
\epsscale{1.0}
\plotone{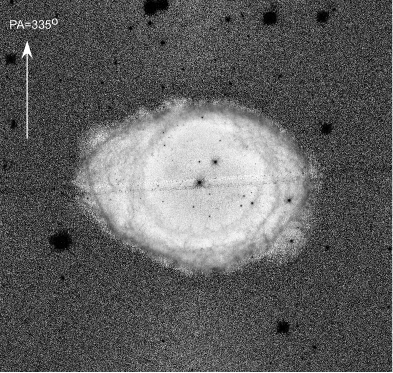}
\caption{This image of the derived Equivalent Width of the \Hb\ emission line is for the same FOV as Figure 3 of the preceding paper. The stitching caused by the dithering across the gap between the two CCD's is even more obvious, but does
not affect the derived equivalent width. The positive Y axis points towards PA = 335\arcdeg. The variations in the Equivalent Width are discussed in the text.
\label{fig:f8}}
\end{figure}

The surprise is that the EW drops  dramatically in the outer parts of the Main Ring. The relative continuum is increasing in strength as the outer ionization front is approached. The same trend is seen in Figure~\ref{fig:f8}, where in addition to the  outer Main Ring drop in EW, there are multiple arcs of lower EW values. Although these resemble the arcs of knots in Figure~\ref{fig:f8}, the effects of small-scale extinction are nulled-out at the two similar wavelengths used in the derivation. The low EW arcs systematically occur within or just outside of the large-scale and small-scale \nii\ features that indicate \nzone\ zones that arise on the Central Star side of ionization fronts. This suggests that the lower EW values are due to light scattered from the dust particles that must be found in the dense PDR's lying just outside of the ionization fronts. This interpretation is strengthened by the fact that the lower EW values in the image coincide with the \htwo\ peaks seen in Figure 3 of the preceding paper.

In order to quantify the amount of excess continuum emission, we
constructed templates from the H\(\beta\) and \ion{He}{2} line images to
model the atomic continuum from H\(^{+}\) and He\(^{++}\),
respectively.\footnote{Since we have no WFC3 image of an He~I line, we
also use H\(\beta\) as an approximate proxy for the He\(^{+}\)
continuum.  This is justified since the He\(^{+}\) continuum is a
small fraction (roughly 5\%) of the H\(^{+}\) continuum and the
ground-based spectra show that \ion{He}{1}/H\(\beta\) varies very little
within the nebula (see Fig.~16).}  The gas temperature (assumed
constant within the nebula) was adjusted in order that the predicted
atomic continuum be equal to the observed continuum in regions of the
nebula with the largest EW).  The resultant value
was 14,000~K, which is significantly higher than the temperatures
determined from the \oiii\ and \nii\ lines. Photoionization models confirm that the highly ionized \hezone\  region should be hotter than regions where helium is singly ionized \citep{ost06} and the high kinetic temperature needed to account for the \Hb\  equivalent width seems reasonable.
The modeled atomic continuum is then subtracted from the observed
continuum to give the residual continuum outside of the inner region where EW is a maximum. 
 This residual emission is found to be very well
correlated with the H\(_2\) 2.12~\(\mu\)m emission (Figure~3, Panel F of the preceding paper), which
supports the view that it is produced in neutral/molecular gas and is due to dust scattering.

The total flux of residual nebular emission at 5470~\AA{} calculated
by this method (after removing all stars in the image and correcting
for an extinction of \cHb\  = 0.15) is  2.45$ \times$
10$^{-14}$ erg s$^{-1}$ cm$^{-2}$ \AA$^{-1}$, which is roughly
25\% of the atomic continuum from the nebula.  For comparison, the
total flux from the nebula's central star is 2.06$\times$10$^{-15}$
erg s$^{-1}$ cm$^{-2}$ \AA$^{-1}$, which is over 10 times
less.  This means that the central star cannot be the source of the
scattered continuum seen in the nebula, and the illumination must come
from outside.  The average intensity of the interstellar radiation
field at this wavelength \citep{black} is 1.3$\times$10$^{-7}$
 erg s$^{-1}$ cm$^{-2}$ sr$^{-1}$ \AA$^{-1}$ and, since the
Ring Nebula is about 170~pc above the Galactic plane, the illumination
is from one hemisphere only, so that the angle-averaged mean
intensity, \(J_{\lambda}\) is half this value.  If the scattering
optical depth of the dust in the nebula is \(\tau < 1\), then the
required value of \(J_{\lambda}\) in order to produce a scattered flux
\(F_{\lambda}\) at the earth is \(J_{\lambda} \simeq F_{\lambda} /
(\tau\Omega)\), where \(\Omega\) is the solid angle subtended by the
nebula.  Taking an upper limit of \(\tau \sim 1\) and an average
radius of \(45''\) for the nebula requires an external radiation field
of J$_{\nu}>$1.6$\times$10$^{-7}$ erg s$^{-1}$ cm$^{-2}$
sr$^{-1}$ \AA$^{-1}$, which is roughly two times larger than expected
for the Galactic diffuse field.  The agreement is fair considering the
uncertainties involved, especially since the Galactic field is
expected to vary from place to place \citep{Parravano:2003}.

\subsection{Small Scale Temperature Variations, the t$\rm^{2}$ Problem and the Abundance Discrepancy Factor}
\label{sec:variations}

There is an elusive problem in gaseous nebulae that has to do with their physical
conditions. The temperatures derived from collisionally excited lines (CL) are higher
than those derived from recombination lines (RL) and from Balmer continua. The
difference is considerably higher than that predicted by sophisticated photoionization
models. Moreover it has important implications for the derivation of chemical
abundance ratios. The abundances derived from collisionally excited lines
assuming thermal homogeneity are typically about a factor of two to three times
smaller than those derived from recombination lines.
To reconcile both types of abundances it is necessary to invoke the presence of temperature
variations considerably higher than those predicted by photoionization models.

We address the problem of finding the mechanism or mechanisms responsible for these
temperature variations by studying the distribution of the temperature across the face of \ngc\ based on the very high spatial resolution provided by the HST observations.
The filters available permit us to determine highly accurate T(O$^{++}$) and T(N$^{+}$) values.
There are two observational quantities that are most useful:
the determination of the mean square temperature variation in the plane
of the sky and the study of the typical spatial size of the temperature variations.
In order to obtain the observed mean square temperature variation in the plane of the
sky we will follow the set of equations introduced by Peimbert (1967)
and O'Dell et al. (2003).

The most meaningful measure of small scale temperature variations is the volumetric variation, for which we define the pair of equations
for the average temperature, $T_0(X^{+i})$, and the mean square
temperature fluctuation, $t^2(X^{+i})$,  given by 

\begin{equation}
T_0(X^{+i}) = 
\frac{\int T_e({\bf r}) n_e({\bf r}) n(X^{+i};{\bf r}) dV}
{\int n_e({\bf r}) n(X^{+i};{\bf r}) dV},
\end{equation}
 
 and 

\begin{equation}
t^2(X^{+i}) = 
\frac{\int (T_e - T_0(X^{+i}))^2 n_e n(X^{+i}) dV}
{T_0^2 \int n_e n(X^{+i}) dV},
\end{equation}

respectively, where $T_e$ and $n_e$ are the local electron temperature and density, $n(X^{+i})$
is the local ion density corresponding to the observed emission line,
and $V$ is the observed volume.

However, our observational results are limited to determination of  variations in columns across the plane of the sky \tsqa (O$^{++}$) and \tsqa (N$^{+}$).  This requires the definition of a columnar temperature $T_c$



\begin{equation}
T_c(X^{+i}; \alpha, \delta) = 
\frac{\int  T_e n_e n(X^{+i}) dl}{\int n_e n(X^{+i}) dl},
\end{equation}

where $\alpha$ is the right ascension and $\delta$ is the declination
of a given line of sight (corresponding to a given pixel). Then, the average temperature 
can be written as

\begin{equation}
T_0(X^{+i}) = 
\frac{\int T_c(X^{+i}; \alpha, \delta)\int n_e n(X^{+i}) dl dA}
{\int\int n_e n(X^{+i}) dl dA},
\end{equation}

and the mean square temperature variation over the plane of the sky,
$t^2_A(X^{+i}$), can be defined as:

\begin{equation}
t^2_A(X^{+i}) = 
\frac{\int (T_c(X^{+i})- T_0(X^{+i}))^2 \int n_e n(X^{+i}) dl dA}
{T_0(X^{+i})^2 \int \int n_e n(X^{+i}) dl dA}.
\end{equation}

\subsubsection{Determination of $t^{2}_A$(O$^{++}$), and $t^{2}_A$(N$^{+}$) from the WFC3 images}
\label{sec:Will}
 In order to isolate the effects of the small-scale temperature fluctuations in the plane of the sky, we first ``detrend'' the temperatures by removing the systematic radial variation.
 For \oiii{} we adopted a mean temperature that varies with projected radius, \(R\),
from the central star as
\begin{equation}
  \label{eq:tfit}
  T_0 (\mathrm{O^{++}}, R) = \frac12 \bigl( T_{\mathrm{in}} +  T_{\mathrm{out}} \bigr)
  + \frac12 \bigl( T_{\mathrm{out}} -  T_{\mathrm{in}} \bigr)
  \tanh\left( \frac{ R - R_0 }{\delta R} \right) ,
\end{equation}
with parameters determined from fitting to the observed temperatures:
inner temperature \(T_{\mathrm{in}} = 11,350\)~K; 
outer temperature \(T_{\mathrm{out}} = 9560\)~K; 
transition radius \(R_0 = 17.6''\); 
transition width \(\delta R = 4.4''\).
In the case of \nii{}, no clear radial trend is visible (Figure~\ref{fig:f3} and Figure~\ref{fig:f4}) 
so a constant mean temperature was used. 
The uncertainties in the pixel-to-pixel temperature measurements 
were dominated by uncertainty in the measurement of the auroral lines,
\nii ~575.5~nm and \oiii ~436.3~nm,
since these were roughly 100 times fainter than the corresponding nebular lines. 
The principal contributions to these uncertainties were 
(1) shot noise associated with the Poisson statistics of the arriving photons; 
(2) readout plus dark noise associated with the CCD; 
(3) pixel-to-pixel flat-field variations; 
(4) cosmic rays artifacts. 
Noise sources (2) and (3) are very well understood and characterized 
for the WFC3 instrument,
and their effect on the measured \(t^2\) was found to be negligible. 
Source (1) is also well understood and will produce a contribution
to the observed fluctuations of \(t^2_{\mathrm{noise}} = 1 / (s^2 N)\), 
where \(N\) is the number of detected photoelectrons 
and \(s \equiv d \ln T / d \ln r\) is the sensitivity of the derived temperatures 
to the diagnostic line ratio \(r\)
(at \(10^4\)~K \(s = 2.5\) for \nii{} and \(s = 3.29\) for \oiii{}).  
Source (4) was the most problematic, 
since 5--9\% of the pixels in each WFC3 exposure are affected by cosmic ray artifacts
\citep{raj10},
and considerable care had to be taken taken to minimize its effects on our analysis.

Multiple exposures were taken of the auroral lines 
(3 exposures of 900~s each for the FQ436N filter,
4 exposures of 600~s each for the FQ575N filter),
which were combined in the standard pipeline reduction process that carries out
automated cosmic ray artifact rejection
\citep{bus10}.
Although the automated pipeline process 
effectively removes the bright cores of cosmic ray artifacts, 
many cosmic ray artifacts possess fainter halos that are still visible on the combined images.
We therefore employ two filters in order to restrict our \(t^2\) analysis 
to only the highest quality portions of the image.
First, we rejected all pixels which were flagged as bad by the pipeline process 
in even one exposure.
Second, we rejected all pixels in which the standard deviation 
between the three (or four) individual exposures was larger than 
a certain factor times the expected width of the Poisson distribution, 
\(\sqrt{N}\).
We found that a value of 1.5 for this factor is a good compromise between 
effectively removing low-level cosmic ray artifacts, 
while at the same time not throwing away too many pixels. 
The fraction of image pixels that are rejected by these two filters was 30 to 40\%. 

In order to both 
reduce the effects of Poisson noise and
investigate the spatial scale of the plane-of-sky temperature fluctuations,
we calculated \(t^2_{\mathrm{A}}\) for a series of \(m \times m\) binnings of the images: 
\(1 \times 1\) pixel, \(2 \times 2\) pixels, \(4 \times 4\) pixels, etc., 
up to \(128 \times 128\) pixels. 
The binning was performed by averaging the surface brightnesses 
of the auroral and nebular lines, 
but for only the good pixels (defined as in  previous paragraph) 
in each \(m \times m\) patch. 
The mean temperature of each patch was calculated as in \S~3.3 and 
\(t^2_{\mathrm{A}}\) was calculated as in Equation~(7), 
but using the surface brightness of the nebular line as a proxy for
\(\int n_{\mathrm{e}}\, n(\mathrm{X}^{+i}) \,dl\). 
At the same time, the sum of the total number of detected photoelectrons 
that contribute to each patch was tracked,
so that \(t^2_{\mathrm{noise}}\) could be calculated as above. 

\begin{figure}
  \setkeys{Gin}{height=0.48\linewidth}
  \setlength\tabcolsep{1em}
  \begin{tabular}{@{}ll@{}}
    (a) & (b) \\
    \includegraphics{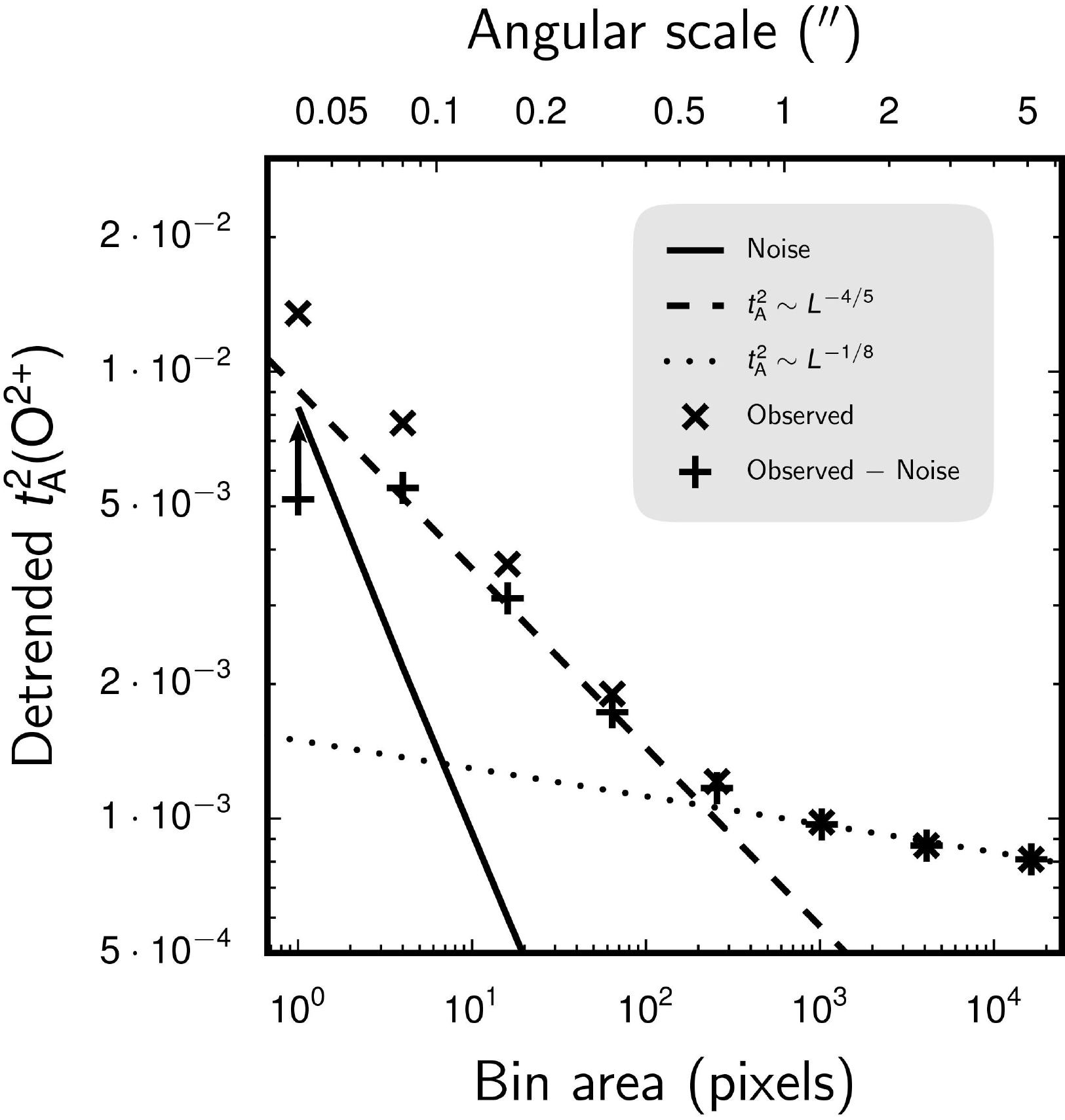} &
    \includegraphics{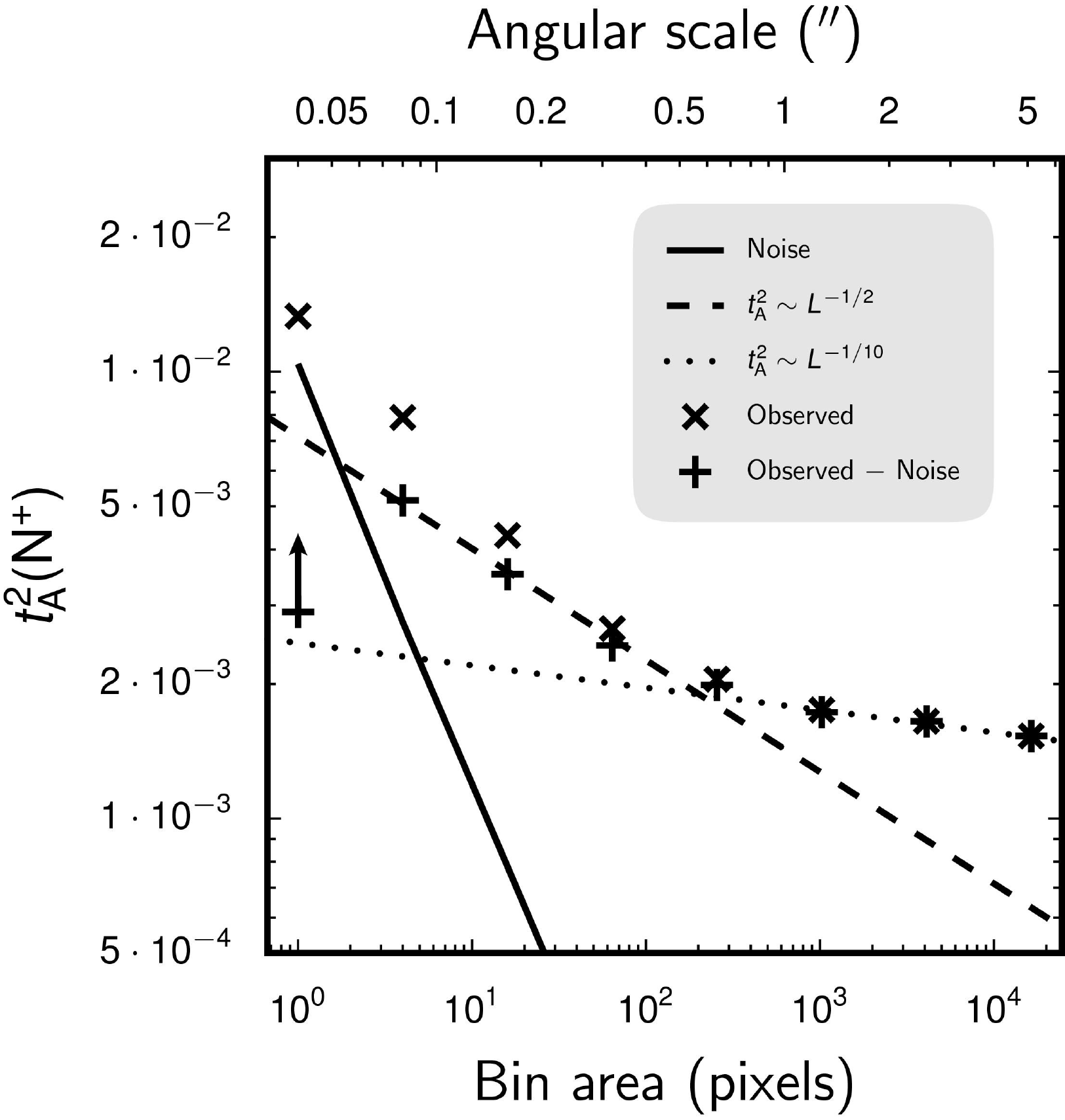} \\
  \end{tabular}
  \caption {
    Global plane-of-sky temperature fluctuations \(t^2_{\mathrm{A}}\) for  (a) \oiii{} and (b) \nii{} as a function of bin area (lower horizontal axis) 
    or angular scale (upper horizontal axis). 
    Cross symbols show the measured values, 
    while plus symbols show the values after correcting for the expected contribution 
    from noise (indicated by the solid line). 
    The left-most point in each graph is affected 
    by the finite angular resolution of the telescope 
    and hence is only a lower limit to the true \(t^2_{\mathrm{A}}\) at that scale.
   In the case of \oiii, ``detrended'' temperatures are used, from which the large-scale radial variation has been removed as described in the text.
  \label{fig:f9}}
	\end{figure}

The results are shown in Figure~\ref{fig:f9} for \oiii{} and \nii{} respectively. 
The cross symbols show the measured \(t^2_{\mathrm{A}}\), 
while the solid line shows the expected contribution from noise, 
which, in addition to Poisson and read-out noise, 
also includes the a very small (\(< 10^{-4}\)) contribution 
from flat-field variations. 
The plus symbols are the result of subtracting the expected noise contribution
from the observed values. 
It should be noted that the diffraction-limit resolution of the images 
is about 2 pixels and so the results of binning at \(1 \times 1\) pixels
should be considered lower limits on the true fluctuations at that scale. 

The noise-corrected results for both ions can be well fit by 
piecewise power laws of the form \(t^2_{\mathrm{A}} \propto L^a\), 
where \(L\) is the angular scale, 
with a break in the power index \(a\) at a scale of \(L_{\mathrm{break}} \simeq 0.5''\) as shown in Table~1. 
The fall off of the fluctuations with scale is significantly steeper
at scales smaller than \(L_{\mathrm{break}}\) 
(\(a \simeq -\frac45\) for \oiii{} and \(a \simeq -\frac12\) for \nii{})
than at larger scales 
(\(a \simeq -\frac18\) for \oiii{} and \(a \simeq -\frac{1}{10}\) for \nii{}).

\subsubsection{Determination of the $t^2$ volumetric factor for selected ions}
\label{sec:tsqobs}
Up to this point we have presented the results in terms of $t^2_A$(O$^{++}$) and $t^2_A$(N$^{+}$), but
to compare with photoionization models and with observations of the whole object
what is needed are the variations in three dimensions $t^2$($X^{+i}$).
In order to obtain the total $t^2$($X^{+i}$)values we need to consider the
variations along the line of sight. It can be shown that the relevant
equation is:

\begin{equation}
t^2(X^{+i}) = t^2_A(X^{+i}) + \left<t^2_c(X^{+i}; \alpha, \delta)\right>,
\end{equation}

where $t^2_c$, the variation along a given line of sight, is given by:

\begin{equation}
t^2_c(X^{+i}; \alpha, \delta) = \frac{\int n_e n(X^{+i}) (T_e - T_c)^2
dl} {T_c^2 \int n_e n(X^{+i}) dl},
\end{equation}

and the average over all lines of sight is given by:

\begin{equation}
\left<t^2_c(X^{+i}; \alpha, \delta)\right> =
\frac{\int t^2_c T^2_c \int n_e n(X^{+i}) dl dA}
{T_0^2 \int \int n_e n(X^{+i}) dl dA}.
\end{equation}

Liu et al. (2004a,b) have presented slit spectra line intensity observations of \ngc\ 
that give sample  regions across the whole object, while our HST 
observations include approximately one quadrant. We consider our highly sampled quadrant to be 
representative of the whole object. From the data of Liu et al. 
it is found that O is distributed as follows: 32.4\% as O$^{+}$, 52.3\% as O$^{++}$ and 15.3\% 
of O in higher degrees of ionization. The last number comes from the assumption 
that  the ratio of abundance of the higher degrees of ionization to {\it n}(O$^{+}$ + O$^{++}$)
is given by {\it n}(He$^{++}$)/{\it n}(He$^{+}$ + He$^{++}$) and that the He/H abundance ratio 
is 0.112.

We can make three estimates of $T_0$(O$^{++}$) and $t^{2}$(O$^{++}$) from the data by 
Liu et al.(2004a,b). a) By assuming that the nebula is chemically homogeneous and 
that the O~II RL and the \oiii\  CL  originate in the same volume we can obtain 
two temperatures for the O$^{++}$ region, one given by the \oiii\  4363/4959 ratio 
and the other by the O III 4959 to the O II V1 multiplet ratio (these two ratios 
have very different temperature dependences) and by using the formalism presented 
by \citep{pei67}, and the computations by Storey (1994), \citep{bas06} and Storey (see Liu (2012)) 
we have determined the $T_0$ and $t^2$ values 
presented in Table 2. 
b) By assuming that the average temperature and the 
temperature variations are similar in the O$^{+}$ and O$^{++}$ regions we can combine  
$T$(4363/4959) with $T$(He I,5876/7281) and obtain the $T_0$ and $t^2$ values 
presented in Table 2. 
 c) By assuming that $T$(4363/4959) is representative 
of the whole object we can compare it with T(H11/Balmer Jump) and obtain the  
$T_0=$ and $t^2$ values in Table 2.
The three $t^2$ values presented in Table 2 are in good agreement and a
representative $t^2($O$^{++}$) value is in the 0.045 to 0.055 range.

It is not possible to derive $t^2$ and $T_0$ combining $T$(HeI) and $T$(H11/Balmer 
Jump) because the temperature dependence of both determinations is almost the 
same.

An absolute lower limit of the $t^{2}_A$(O$^{++}$) value derived in \ref{sec:Will} is 
0.009, this result is obtained by sampling all the observed area with bins of 2$\times$2 pixels.
The true value of $t^{2}_A$(O$^{++}$) could only be obtained with infinitesimally small sampling.
There is a clear trend of increasing $t^{2}_A$(O$^{++}$) values with smaller linear scale, suggesting
that the true value is higher; but clearly this trend cannot continue to arbitrarily large values. By
extrapolating the $t^{2}_A$(O$^{++}$) values for 8$\times$8, 4$\times$4 and 2$\times$2 pixels to a sampling
area of 1$\times$1 pixel a $t^{2}_A$(O$^{++}$) value of 0.013 is obtained (see Figure~\ref{fig:f9}). This value
should be considered as a soft lower limit to the true $t^{2}_A$(O$^{++}$) value since the value
could be higher if this trend continues to a sampling with areas smaller than 1$\times$1 pixel.
The 0.013 value is about a factor of four smaller than the volumetric $t^{2}$(O$^{++}$)value (see Table 2).
This difference is expected, since the temperatures for each point in the plane of the sky are an
average of the temperatures, see equation (5) and equation (10). As such $t^{2}_A$(O$^{++}$) is only
a lower limit to $t^{2}$(O$^{++}$). The $t^{2}_A/t^{2}$ ratio is also expected to be small
when the characteristic scale of the spatial variations is much smaller than the whole object. 

From geometrical considerations one would 
expect the variations of one additional dimension, $t^2_c$, to be at 
least half as large as the variations of two dimensions, $t^2_A$; 
this works for large scale temperature variations. The power included in $t^2_c$ 
increases in the presence of small size scale variations. This is because the 
many independent thermal elements along each line of sight will be masked 
in the averaging process, greatly reducing the total $t^2_A$ compared to $t^2$.

Similarly, an absolute lower limit of $t^{2}_A$(N$^{+}$) value derived in the previous
subsubsection is 0.0053. This result is obtained by sampling all the observed area with bins
of 2$\times$2 pixels. In this case the trend, showing larger inhomogeneities at small scales, is also
visible. Compared with the $t^{2}_A$(O$^{++}$) determinations the $t^{2}_A$(N$^{+}$)
shows more power at scales larger than 8$\times$8 pixels and a slightly shallower behaviour at
smaller scales. As is the case for O$^{++}$ the "most important scale" is probably smaller than
our detection threshold. By extrapolating the $t^{2}_A$(N$^{+}$) values for 8$\times$8, 4$\times$4 and
2$\times$2 pixels to a sampling area of 1$\times$1 pixel, a $t^{2}_A$(N$^{+}$) value of 0.008 is obtained
(see Figure~\ref{fig:f9}).

The previous discussion implies that the temperature variations are dominated by variations 
in small distance scales of the order of a thousandth of a parsec or less.
At the adopted distance of \ngc\  the length of two pixels amounts to 
about 9$\times$10$^{14}$ cm.      

For comparison O'Dell et al.(2003) obtained for the Orion Nebula $t^{2}_A$(O$^{++}$) = 0.0079 and  $t^2$(O$^{++}$) = 0.021 $\pm$0.005, a ratio of $t^2$(O$^{++}$) over 
$t^{2}_A$(O$^{++}$) of about 2.0--3.3.

\subsubsection {The O/H abundance discrepancy factor}
\label{sec:abundproblem}

New Location as first paragraph of this section. The recombination line abundances of H, He, C, N, O, and Ne have a similar dependence on the electron 
temperature, therefore in the case of chemical homogeneity the abundances derived from the ratio
of two RL are almost independent of the temperature structure. This is not the case for the abundances
of collisionally excited lines of C, N, O, and Ne relative to H because the first four do depend
strongly on the temperature structure. Their intensity increases with temperature; while the opposite
occurs with the recombination lines. In addition, their intensity decreases mildly with increasing temperature. 
Consequently in the presence of temperature variations the preferred abundances are those 
derived from the ratio of two recombination lines. However, an accurate derivation requires measurement of inherently weak and some-times blended lines. This demands high signal to noise spectra of excellent wavelength resolution.

In Table 3 we present the chemical abundances derived by Liu et al. (2004a,b) for \ngc\ based on RL  and on CL lines. The total O abundances are 
smaller by 0.04 dex than those presented by Liu et al. (2004b) because they
did not correct properly for the fraction of O$^{+3}$ based on the He$^{+}$ and He$^{++}$  abundances.
From Table 3 it can be seen that the recombination abundances for C, N, O, and Ne are 
higher than the collisional abundances by about 0.4 dex.
 
There are in the literature two families of results to explain the CL versus the RL abundance 
differences based on the temperature dependence of the abundance determinations: a) the presence of 
low temperature H-poor inclusions that are rich in helium and heavy elements in planetary nebulae 
(e. g. Liu 2006 and references therein), or b) the presence of temperature variations in a
homogeneous medium due to additional physical  processes not included in photoionization 
models (e. g. Peimbert and Peimbert 2006, and references therein). The fact that, $T$(H), 
$T$(He) and $T$(V1/4959) are similar indicates that an important chemical inhomogeneity is not present 
in \ngc. We are left with the idea of a well mixed material with temperature fluctuations. 
The details of the origin of these fluctuations remain unexplained. Physical
mechanisms known to play at least a small part in explaining these fluctuations are: shocks,
magnetic reconnection and shadowed regions, the relative importance of these processes must be studied further.


\subsubsection {Photoionization model predictions}
\label{sec:tsqmodel}

We computed a simple model of the main H~II region of \ngc\ using reasonable parameters.  These included a central star temperature of 1.2$\times$10$^5$ K, a luminosity of 200 L$_{\sun}$, and abundances typical of \ngc\ as determined from collisionally excited lines.  This model reproduced the overall ionization structure and the intensities of the stronger lines.  We then determined the temperature variations across the model, using the methods described in Kingdon \&\ Ferland (1995).  The ``structural t$^2$", the mean variation in temperature across various emission zones, was 0.0062 for the H~II region, and 0.0026 across the \oiii -forming region. This model also predicts for the He$^{++}$ emitting zone that \Te =12500 K, essentially in agreement with the value of 13000 K inferred from the EW in the central region (\S \ref{sec:EWs}).

From observations (\S ~\ref{sec:tsqobs} we have determined
that $t^{2}$(O$^{++}$) = 0.050$\pm$0.008, and $t^{2}_A$(O$^{++}$)$> $0.0090. These
values are considerably higher than the $t^{2}$(O$^{++}$) = 0.0026 value predicted
by the photoionization model. These two results imply that in addition to photoionization there must be one
or more additional physical processes responsible for the observed t$^2$ values. 
Unfortunately, although the relevant scales for these processes ($\le$ 0.5'') are in principle accessible to \textit{HST} observations, the low signal-to-noise of the auroral line images are a serious impediment to determining their origin.

\subsubsection{Possible origins of small-scale temperature fluctuations}
\label{sec:origins}

In efforts to explain the existence of small-scale temperature
fluctuations in photoionized nebulae, various physical mechanisms have
been proposed in the literature.  These mechanisms can be divided into
two classes: first, those that posit an extra source of energy in
addition to the photoionization heating that is generally taken to
dominate the heating within the nebula, and, second, those that
attempt to modify the heating/cooling balance through other means.
 
An example of the the first class of mechanisms is the action of
highly supersonic stellar winds, either directly through the
thermalization of the wind kinetic energy \citep{lur01}, or
via conduction fronts at the boundaries of hot, shocked bubbles
\citep{mac96}.  In the case of \ion{H}{2} regions,
quantitative studies have suggested that these mechanisms are unable
to satisfy the derived energy requirements \citep{bin00}, but
we are unaware of comparable calculations tailored to the case of Planetary Nebulae.
A more promising source of energy is perhaps internal turbulent
motions driven by transonic photoevaporation flows from dense
condensations at the periphery of the nebula (such as the knots
observed in the Ring Nebula, see  \S~2.2.1 of the preceding paper, and where the dissipation
of the turbulent energy would be principally via low Mach number
shocks.  In the case of \ion{H}{2} regions, simulations indicate that
the total kinetic energy of such turbulent motions is similar to the
thermal energy of the ionized gas \citep{art11}, and similar
processes are expected to occur in Planetary Nebulae  \citep{gar06}.
Another potential source of extra heating is magnetic reconnection
\cite{laz99,laz12}, although no quantitative estimate
of its importance in the context of Planetary Nebulae has yet been made.

An example of the second class of mechanism is the possible existence of small-scale pockets of high-metallicity gas \citep{kandf98,rob05,liu06},  which, due to vastly enhanced collisional metal line cooling, will have a lower equilibrium temperature than the rest of the nebula. 
The origin of these pockets may be solid bodies such as comets, which were evaporated during the asymptotic giant branch phase that preceded the formation of the planetary nebula \citep{hen10}.  Another example of this class of mechanism is the existence of diffusely illuminated regions of the nebula that are shadowed from the direct stellar radiation \citep{ode03}. 
These too would be lower temperature than the general nebula, due to being illuminated by the relatively soft ionizing field characteristic of recombination radiation.

\section{Summary and Conclusions}


We have used the flux-calibrated HST WFC3 emission-line images described in the preceding paper to determine the physical conditions in \ngc\  down to sub-arcsec scales. Our major conclusions are listed below.

1. The equivalent width of the continuum using the \Hb\ emission-line as a reference is unexpectedly small, indicating a strong relative continuum. In the inner regions of the nebula this is probably due to the electron temperature being as high as 13000 K and in the outer parts of the nebula it probably indicates scattering of starlight in the diffuse Galactic radiation field, although a value twice the global average of the diffuse Galactic radiation field is required for this interpretation to apply. 

2. The electron temperature as determined from nebula to auroral transitions of singly ionized nitrogen \nii\ and doubly ionized oxygen \oiii\ varies systematically. \Te\ increases from at or slightly lower than 10000 K in the outer region (from \nii) to about 11500 K in the inner-most region (from \oiii). Near the central star there is a region of \hezone, where oxygen will be triply ionized. This means that lines of sight near the central star measure material in the Lobes and the higher temperature derived from the equivalent width is characteristic of the innermost region of the nebula.

3. About half of the power in the observed
  t$^2_\mathrm{A}(\mathrm{O}^{++}$) and
  t$^2_\mathrm{A}(\mathrm{N}^{+}$) values originates in structures
  smaller than 8~pixels (\(3.5 \times 10^{15}\)~cm). 
  After correcting
  for the presence of structures smaller than \(2 \times 2\) pixels as
  well as the contribution of the variation of the temperature along
  the line of sight, it follows that most of the volumetric \(t^2\)
  comes from structures with typical sizes of 2$\times$2 pixels
  (\(9 \times 10^{14}\)~cm). 
  These small scales suggest the presence
  of one or more physical processes,
  small scale chemical inhomogeneities, radiation shadowing,
magnetic reconnection, or dissipation of turbulent energy in shocks as possible causes
for the large \(t^2\) values.
  
 \acknowledgments
We are grateful to David Thompson of the Large Binocular Telescope Observatory for providing copies of his unpublished LBTO \htwo\ data taken with the LUCI1 instrument and
to Antonio Peimbert for several fruitful discussions. 

GJF acknowledges support by NSF (0908877; 1108928; and 1109061), NASA (10-ATP10-0053, 10-ADAP10-0073, and NNX12AH73G), JPL (RSA No 1430426), and STScI (HST-AR-12125.01, GO-12560, and HST-GO-12309).  MP received partial support from CONACyT grant 129553. WJH acknowledges financial support from DGAPA--UNAM through project PAPIIT IN102012. CRO's participation was supported in part by HST program GO 12309. 

{\it Facilities:} \facility{HST {(WFC3)}}


\clearpage



\begin{deluxetable}{lcc}
\tabletypesize{\scriptsize}
\tablecaption{Contributions to noise-corrected plane-of-sky temperature fluctuations}
\label{tab:t2contributions}
\tablewidth{0pt}
\tablehead{
\colhead{Source} &
\colhead{$t^{2}_{A}$(\oiii)} &
\colhead{$t^{2}_{A}$(\nii)}}
\startdata
Large-scale radial gradient (\(L > 5''\))                           & 0.004 &  $\sim$ 0  \\
Fluctuations on medium scales (\(L = 0.5''\) to \(5''\))   & 0.001 ($\propto$ L$^{-1/8}$)) & 0.002~($\propto$ L$^{-1/10}$) \\
Fluctuations on small scales (\(L = 0.08''\) to \(0.5''\))  & 0.004 ($\propto$ L$^{-4/5}$)) & 0.003~($\propto$ L$^{-1/2}$) \\ 
\hline
Total nebular plane-of-sky fluctuations (\(L > 0.08''\)) & 0.009 & 0.005 \\
 \enddata
 \end{deluxetable}

\begin{deluxetable}{lcc}
\tabletypesize{\scriptsize}
\tablecaption{$T_0$  and $t^2$ Determined from Slit Spectroscopy*}
\label{tab:tresults}
\tablewidth{0pt}
\tablehead{
\colhead{$T_0$} &
\colhead{$t^2$} &
\colhead{Temperatures Used}}
\startdata
8775$\pm$270 &  0.057$\pm$0.008   & $T$(4363/4959) = 10630 and  $T$(V1/4959) = 8810 \\
9050$\pm$370 &  0.052$\pm$0.011   & $T$(4363/4959) = 10630 and  $T$(He I, 5876/7281) = 8290 \\
9630$\pm$750  & 0.032$\pm$0.022   & $T$(4363/4959) = 10630 and  $T$(H11/Balmer Jump) = 9100 \\
\enddata
\tablecomments{~*The line and continuum fluxes are from the entire nebula as described in the text.}
\end{deluxetable}

\begin{deluxetable}{lcccccc}
\tabletypesize{\scriptsize}
\tablecaption{The Ring Nebula Abundances Determined from Slit Spectroscopy}
\label{tab:abundances}
\tablewidth{0pt}
\tablehead{
\colhead{} &
\colhead{H} &
\colhead{He} &
\colhead{C} &
\colhead{N} &
\colhead{O} &
\colhead{Ne}}
\startdata
RL & 12.00 & 11.05 & 9.10 & 8.62 & 9.14 & 8.71 \\
CL &    ...     &   ...      &  8.59 & 8.22 & 8.76 & 8.23\\
\enddata
\tablecomments{All abundances are in base 10 logarithms relative to the abundance of H set to 12.00}
\end{deluxetable}

\end{document}